\documentclass[a4paper,fleqn]{cas-dc}

\usepackage[numbers]{natbib}
\usepackage{tabularx}
\newdefinition{rmk}{Definition}
\graphicspath{{figs/}}
\usepackage{graphicx}
\usepackage{subcaption}
\usepackage{threeparttable}
\usepackage{algorithmic}
\usepackage{algorithm}
\usepackage{pifont}
\def\tsc#1{\csdef{#1}{\textsc{\lowercase{#1}}\xspace}}
\tsc{WGM}
\tsc{QE}

\begin{document}
\let\WriteBookmarks\relax
\def\floatpagepagefraction{1}
\def\textpagefraction{.001}

\title [mode = title]{CLIProv: A Contrastive Log-to-Intelligence Multimodal Approach for Threat Detection and Provenance Analysis}  


\tnotetext[1]{<This work was supported by the Natural Science Foundation of China under Grant [U21B2020]; Fundamental Research Funds for the Central Universities (Beijing University of Posts and Telecommunications) for Action Plan under Grant [2021XD-A11-3]>.} 

%

\author[1]{Jingwen Li}[style=chinese]
\credit{Conceptualization, Methodology, Writing–original draft}

\affiliation[1]{organization=Beijing University of Posts and Telecommunications,
            city=Beijing,
            postcode=100876, 
            country=China}
            
\author[1]{Ru Zhang}[style=chinese, orcid=0000-0001-6641-3236]
\cormark[1]
\ead{zhangru@bupt.edu.cn}
\credit{Supervision, Writing-Review $\&$ Editing}            
            
\author[1]{Jianyi Liu}[style=chinese]
\credit{Methodology, Writing-Review $\&$ Editing, Resources}

\author[2]{WanGuo Zhao}[style=chinese]
\credit{Data curation, Resources}

\affiliation[2]{organization=Beijing Anheng Xin'an Technology Co., Ltd,
            city=Beijing,
            postcode=100089, 
            country=China}
            
\cortext[1]{Corresponding author}



\begin{abstract}
With the increasing complexity of cyberattacks, the proactive and forward-looking nature of threat intelligence has become more crucial for threat detection and provenance analysis. However, translating high-level attack patterns described in Tactics, Techniques, and Procedures (TTP) intelligence into actionable security policies remains a significant challenge. This challenge arises from the semantic gap between high-level threat intelligence and low-level provenance log. To address this issue, this paper introduces CLIProv, a novel approach for detecting threat behaviors in a host system. CLIProv employs a multimodal framework that leverages contrastive learning to align the semantics of provenance logs with threat intelligence, effectively correlating system intrusion activities with attack patterns. Furthermore, CLIProv formulates threat detection as a semantic search problem, identifying attack behaviors by searching for threat intelligence that is most semantically similar to the log sequence. By leveraging attack pattern information in threat intelligence, CLIProv identifies TTPs and generates complete and concise attack scenarios. Experimental evaluations on standard datasets show that CLIProv effectively identifies attack behaviors in system provenance logs, offering valuable references for potential techniques. Compared to state-of-the-art methods, CLIProv achieves higher precision and significantly improved detection efficiency.
\end{abstract}




\begin{keywords}
Threat intelligence\sep Threat detection\sep Provenance analysis\sep Multimodal\sep Contrastive learning\sep Attack scenarios reconstruction
\end{keywords}
\maketitle

\section{Introduction}\label{sec:1}
Advanced Persistent Threats (APTs), characterized by advanced techniques, long duration, and high targeting, have become one of the most significant threats in cybersecurity \cite{math11143115}. APT attackers typically gain initial access by exploiting zero-day vulnerabilities \cite{3530812} or employing social engineering techniques \cite{10273652}. Then, they establish persistence and perform lateral movement to remain hidden in the system for the long term, gradually infiltrating critical networks. These activities severely impact critical industries such as energy and telecommunications, and even threaten national security and public safety. For example, the VPNFilter incident was a notorious attack in 2018 targeting telecommunications devices, leveraging router security vulnerabilities to achieve initial access \cite{yu2024cost}. The US Department of Justice has linked the incident to APT28 \cite{VPNFilter}. The attack compromised more than half a million devices across at least 54 countries and regions. 

Traditional intrusion detection methods based on network traffic can only monitor network-level patterns and fail to capture the behavioral logic behind data packets. Moreover, APT attacks are typically distributed and multi-stage. Traffic-based methods lack the ability to correlate attack behaviors, making it difficult to identify the infection source and track the attack path \cite{10237298}. The widespread use of encryption and obfuscation techniques further exacerbates these limitations, rendering traffic-based approaches insufficient for decrypting and interpreting packet content. According to recent studies \cite{YangXXLZ23, chen2022apt, altinisik2023provg}, provenance logs may be a more reliable data source for identifying APTs. Provenance logs are structured audit logs describing the system execution history \cite{KAIROS}. They include the interaction relationships between system entities, such as the read-and-write interactions between processes and files. The rich contextual relationships contained in provenance logs provide a comprehensive view of system behavior \cite{291066}, offering strong support for threat detection. 

Threat detection methods based on provenance logs are mainly classified into two categories: anomaly detection and threat hunting. Anomaly detection \cite{YangXXLZ23} identifies attacks by monitoring deviations from established normal behavior patterns, using machine learning or statistical models to detect anomalies within the system. Threat hunting \cite{chen2022apt, altinisik2023provg} leverages experience and intelligence to establish reasonable attack hypotheses, which uses knowledge templates to proactively identify signs of attacks. With the development of threat intelligence, query graphs \cite{milajerdi2019poirot} have emerged as a mainstream knowledge template. The query graph extracts Indicators of Compromise (IOCs) and their relationships from threat intelligence, visually representing the interactions between different system entities in the form of a directed graph, which helps identify and correlate complex multi-step attacks. Although current solutions exhibit outstanding detection performance, they still have certain inherent limitations.

Limitation 1. Anomaly detection heavily rely on the quality and representativeness of the training data, which often results in a high rate of false positives. To minimize false positives, the training data should ideally cover all normal behavior patterns, which poses a challenge to computational resources.

Limitation 2. Threat hunting rely on structured query templates. These templates often require manual design, which is time-consuming and labor-intensive. Moreover, the exact matching approach lacks flexibility and struggles to identify unknown attacks.

Limitation 3. The scale of provenance logs limits the efficiency of threat queries. APT attacks are hidden and usually have a long period of time, resulting in a large scale of logs, which brings challenges to provenance analysis. Query operations in large-scale data graphs, such as path searches and pattern matching, require traversing numerous nodes and edges, leading to slow response times.

\begin{figure}[!t]
\centering
\includegraphics[width=3.5in]{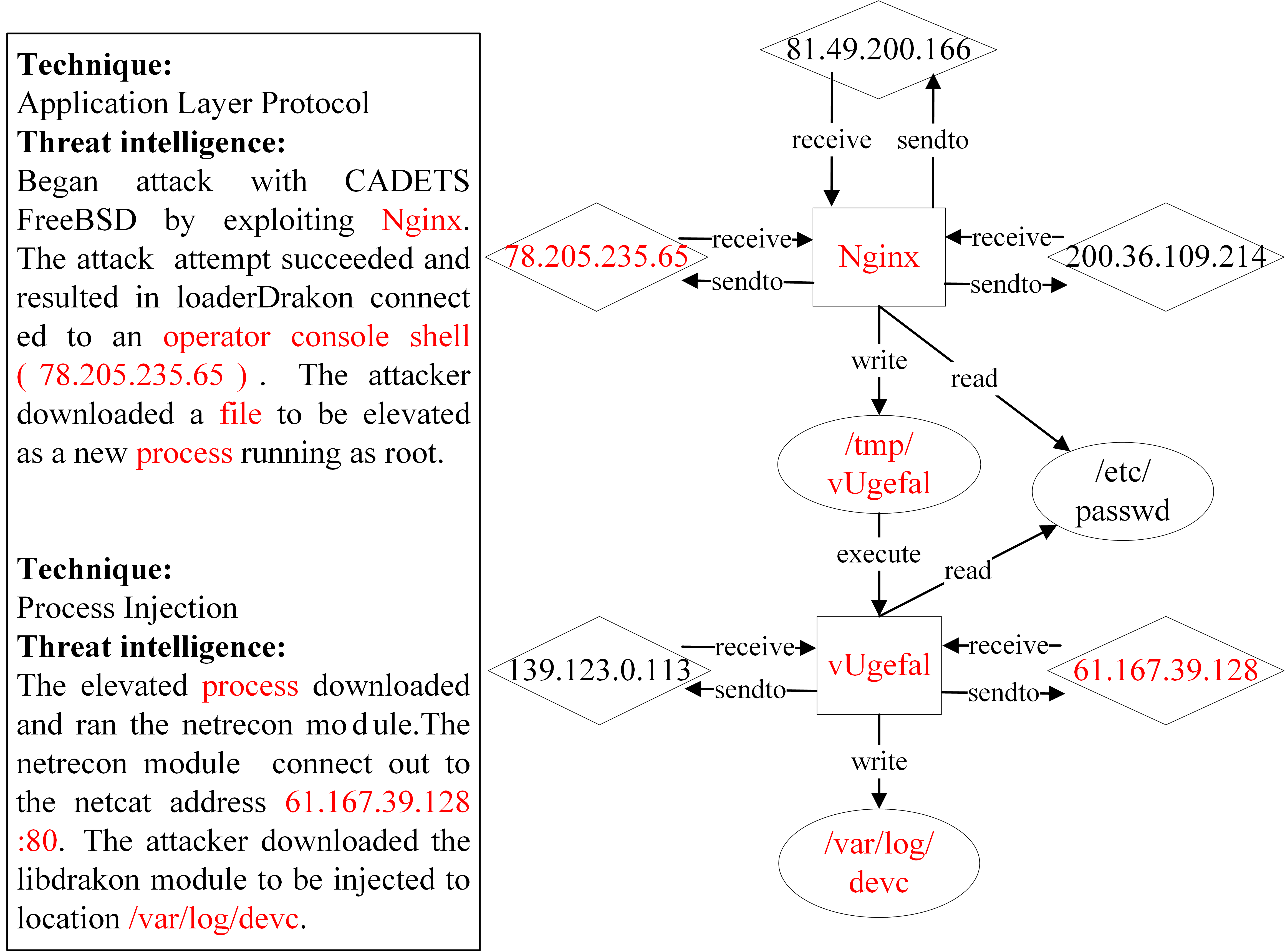}
\caption{An attack scenario example in the DRAPA TC dataset.}
\label{fig_1}
\end{figure}

Based on the above limitations, this paper considers incorporating threat intelligence to enhance the understanding of attack behavior patterns. This is an underexplored area for APT threat hunting. Fig.~\ref{fig_1} shows an attack scenario in the DRAPA TC \cite{transparent} dataset, including the threat intelligence and provenance graph. It highlights the corresponding part of threat intelligence and provenance logs in red. As shown in Fig.~\ref{fig_1}, threat intelligence provides descriptive information of the attack process in natural language form. By learning the correlation between threat intelligence and data provenance, the tactical and technical information in threat intelligence can help gain a deeper understanding of attack behavior semantics. However, the differences in semantic levels and syntactic structures pose challenges in aligning high-level intelligence with low-level logs.

This paper proposes CLIProv, a novel threat hunting method based on multimodal learning. CLIProv focuses on learning attack behavior patterns from provenance logs instead of learning normal behavior patterns, which rely on numerous normal samples. It incorporates threat intelligence to enhance the model's understanding of complex attack semantics and improve its generalization across scenarios. CLIProv directly utilizes natural language threat intelligence and raw provenance logs to construct positive and negative sample pairs, avoiding the need for secondary extraction of intelligence. Through contrastive learning \cite{radford2021learning}, the model can automatically capture deep semantic relationships between attack patterns and system behaviors. CLIProv employs a semantic search mechanism for threat hunting. It divides the provenance logs based on the contextual dependencies of system entities and the density of operational behaviors, then maps the high-dimensional raw data to a low-dimensional vector space, reducing computational complexity and resource consumption.

In summary, the contributions of this paper are as follows.

\begin{enumerate}[(1)]
\item{A representation learning method of provenance logs is proposed. CLIProv is the first framework to apply multimodal techniques to the joint learning of threat intelligence and provenance logs. Using contrastive learning to align the embeddings of logs and intelligence within a shared semantic space, CLIProv generates efficient and consistent representations for different data corresponding to the same attack patterns.}
\item{A threat investigation method based on semantic search is proposed. Without relying on precise query templates, CLIProv accurately identifies attack behaviors by searching for log sequences that match the attack patterns described in threat intelligence within a shared semantic space.}
\item{We evaluated the detection performance of CLIProv on four public datasets. The results demonstrate that CLIProv performs effectively across scenarios, vulnerabilities and datasets, with higher precision and lower computational costs. The reconstructed attack scenario can provide an interpretable, high-level technical summary.}
\end{enumerate}

The remainder of the paper is organized as follows. Section \ref{sec:2} reviews related work on APT threat detection using provenance logs. Section \ref{sec:3} provides the background and motivation for the proposed methods. Section \ref{sec:4} defines the threat scope and research questions. Section \ref{sec:5} describes the framework and methodological design of the CLIProv. Section \ref{sec:6} presents a comprehensive summary of CLIProv’s experimental results. Finally, Sections \ref{sec:7} and \ref{sec:8} discuss the limitations of the proposed method and conclude the paper, respectively.

\section{Related work}\label{sec:2}

Recently, threat detection methods based on provenance logs can be broadly categorized into anomaly detection and threat hunting. In threat hunting methods, threat intelligence has gradually become an essential component of the attack knowledge system. This section provides a detailed review of related work in these areas and summarizes the progress of the research.

\subsection{Threat detection method based on anomaly detection} \label{sec:2-1}
Anomaly detection methods assume that attack behaviors differ significantly from normal behaviors. These methods detect threats by modeling normal behavior patterns. StreamSpot \cite{manzoor2016fast} proposes a graph-level detection approach. It designs a heterogeneous graph similarity function based on the relative frequency of local substructures and identifies threats using a clustering approach. UNICORN \cite{HanP0MS20} addresses the dynamic nature of attack behaviors by introducing summary encoding during graph evolution. However, graph-level feature extraction methods are insufficiently sensitive to small-scale anomalies, especially as APT attacks often conceal their malicious activities within a large volume of normal operations. Log2vec \cite{liu2019log2vec} introduces a log entry-level detection approach by employing an improved random walk algorithm to learn embeddings of log entries. Moreover, Threatrace \cite{wang2022threatrace} proposes a node-level detection. It customizes the Graph Sample and Aggregate (GraphSAGE) framework to capture contextual features of benign nodes and identify anomalies via structural differences from benign nodes. KAIROS \cite{KAIROS} evaluates node anomalies by examining dynamic edge changes and neighborhood structures. Despite their strengths, anomaly-based methods face challenges as the volume of normal behavior patterns increases, requiring continuous updates to maintain performance and effectiveness.

\subsection{Threat detection method based on threat hunting}\label{sec:2-2}

Threat hunting methods use existing or anticipated knowledge of attack behaviors to define heuristic rules and match them against provenance logs for attack detection.

APTHunter \cite{mahmoud2023apthunter} constructs attack queries by defining key elements such as the source subject, target object, syscall details, intermediate process, and the process tree path length. To address the need for real-time threat search and efficient data storage in large systems, ThreatRaptor \cite{gao2021enabling,gao2021system} introduces the Threat Behavior Query Language (TBQL), whose key primitives facilitate efficient querying of complex multi-step system behaviors. MEGR-APT \cite{aly2024megr} optimizes memory usage by proposing an RDF-based provenance graph representation. It extracts candidate subgraphs to disk and performs threat queries in memory, significantly reducing memory consumption. Similarly, Yu et al. \cite{yu2024cost} focus on IoT environments and propose a cost-effective approach for generating snapshots of low-frequency provenance graphs in the spatial dimension to address resource constraints. By applying predefined inference rules, they identified files and sockets directly or indirectly causally related to symptom nodes, constructing comprehensive attack scenarios. These methods leverage precise query rules and efficient graph representations to enhance the speed and accuracy of threat hunting.

Understanding attacker tactics and motivations is essential for detecting APT attacks. Tactics, Techniques, and Procedures (TTPs) \cite{strom2018mitre} provide a widely used framework for categorizing APT strategies and techniques. A high-level summary of attack behaviors helps analysts identify key stages in the attack chain and take proactive measures. Holmes \cite{milajerdi2019holmes} maps log events to suspicious TTPs by defining related event clusters and employs pruning and noise reduction to generate simplified advanced scenario graphs. Building on this, Rapsheet \cite{hassan2020tactical} expands the TTP rules from 16 to 67 and introduces a graph-based threat scoring scheme based on causal relationships in the kill chain. However, these methods heavily rely on manually defined rules, limiting their adaptability. Moreover, exact matching tends to fail when the attack behavior deviates slightly. With the dynamic evolution of attack techniques, these approach faces the challenge of constantly updating the rule library.

\subsection{Threat hunting method based on threat intelligence} \label{sec:2-3}

With the development of threat intelligence, researchers tend to use attack knowledge in intelligence to reduce the dependence on traditional query rules, which improves detection accuracy and enhances adaptability to evolving attack techniques. Poirot \cite{milajerdi2019poirot} first introduced the concept of the query graph, which leverages the causal relationships in threat intelligence. It represents IOCs as nodes and their interrelations as edges, forming a generalized model of attacker behavior. Poirot framed threat hunting as a graph pattern matching (GPM) problem and proposed an approximation function to evaluate the consistency between query and source graphs, effectively addressing the NP-completeness of graph matching. However, structure-based matching methods like Poirot may fail when attack scenarios are fragmented. To overcome this limitation, Deephunter \cite{wei2021deephunter} and ProvG-Searcher \cite{altinisik2023provg} employ Graph Neural Networks (GNNs) to learn the vector representation of provenance graphs, improving query effectiveness significantly. While query graphs improve the robustness of threat hunting, their manual extraction by security experts remains labor-intensive and time-consuming, highlighting the need for automated solutions.

Extractor \cite{satvat2021extractor} applies Natural Language Processing (NLP) techniques to extract threat entities and relationships from threat intelligence and designs a semantic role labeling method to generate query graphs automatically. AttacKG \cite{li2022attackg} maps query graphs to TTPs, enriching them with higher-level semantic features. With the development of cybersecurity knowledge graphs, recent research \cite{kaiser2023attack,kurniawan2022krystal,ren2022cskg4apt} apply graph traversal algorithms and link prediction techniques to generate attack hypotheses.

\section{Background \& motivation} \label{sec:3}

In this section, we introduce the necessary background knowledge for the implementation of CLIProv and provide a motivating example to highlight its differences and advantages over other threat hunting methods based on threat intelligence.

\subsection{Data provenance and threat intelligence}\label{sec:3-1}

\begin{figure*}[!t]
\centering
\includegraphics[width=7in]{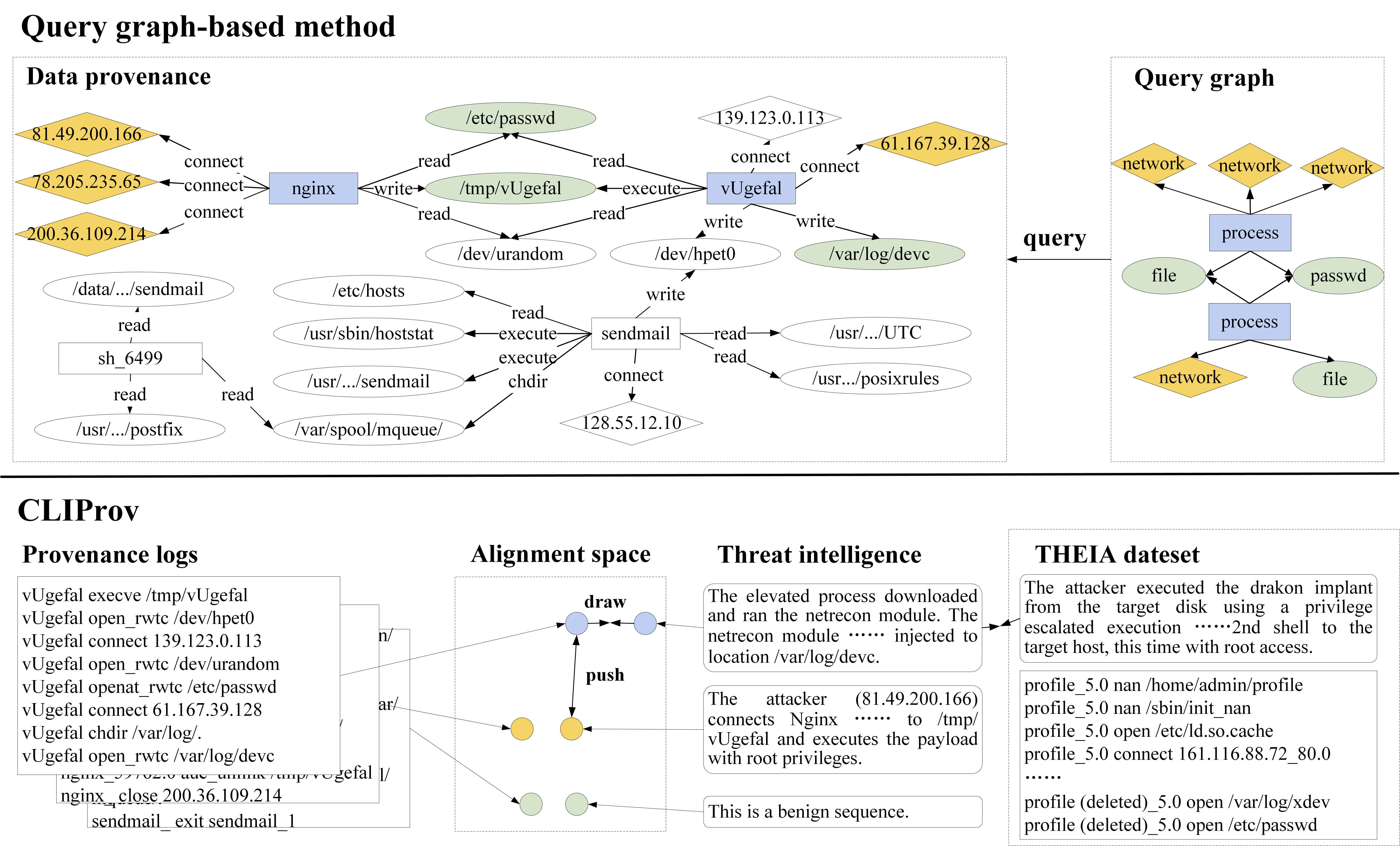}
\caption{Comparison of query graph-based method and CLIProv.}
\label{fig_2}
\end{figure*}

Provenance logs are typically generated by system auditing tools such as Windows ETW and Linux Audit, which record activities such as system calls, file accesses, and user logins in detail. Data provenance is modeled as a directed graph based on the flow of information in these logs, as shown in Fig.~\ref{fig_2}. Nodes represent system-level objects, and edges represent their interactions. 

Threat intelligence is evidence-based knowledge, including context, mechanisms, indicators, implications and actionable advice, about an existing or emerging menace or hazard to assets that can be used to inform decisions regarding the subject's response to that menace or hazard \cite{gartner}. It can provide valuable hypotheses for provenance analysis.

However, data provenance is low-level, micro, and structured information that records actions between system entities. In contrast, threat intelligence is formulated in natural language \cite{sun2023cyber}, providing a high-level, macro perspective on threats, including comprehensive analyses of threat intentions and strategies. There is a semantic gap between them due to their differences in structure and expression. According to the 2024 SANS threat intelligence survey \cite{SANS}, 74\% of threat intelligence is published in the form of reports, but the majority of applied intelligence consists of IOCs. Although threat intelligence in natural language form contains a lot of attack knowledge, a good integration method with threat hunting has not yet been found.

\subsection{Multimodal learning}\label{sec:3-2}

In the field of representation learning, ``modality'' refers to the specific encoding type applied to information \cite{baltruvsaitis2018multimodal}. Multimodal data describe objects from various perspectives, providing complementary information. The feature vectors of different modalities exist in different semantic spaces. Multimodal learning aims to project features from different modalities into a shared subspace, allowing data with similar semantics to have similar feature representations.

To address the semantic gap problem, CLIProv employs multimodal representation alignment method. This approach learns individual representations for each modality while introducing constraints to ensure coordination between them. Given two datasets $X$ and $Y$ of different modalities, the formula for multimodal representation alignment is as follows.

\begin{equation}
\label{eq_1}
f(x;{{W}_{f}})\sim g(y;{{W}_{g}})
\end{equation}
                                 
Where $f$ and $g$ are projection functions that map the original spaces into a shared multimodal alignment space. $\sim$ represents coordination methods, including minimizing cosine distance \cite{frome2013devise}, maximizing correlation \cite{andrew2013deep} and so on. Taking the L2 norm as an example, the minimization of the distance between different modal data can be represented as:

\begin{equation}
\label{eq_2}
\underset{\theta }{\mathop{\min }}\,\left\| f(x;{{W}_{f}})-g(y;{{W}_{g}}) \right\|_{2}^{2}
\end{equation}

In multimodal tasks, unsupervised learning can automatically discover the underlying structure and patterns in data through its statistical characteristics and distributions. It can extract more effective feature representations from the data by identifying similarities and differences within the data. Contrastive learning offers a flexible definition of positive and negative samples and demonstrates exceptional performance, making it an outstanding research direction in unsupervised learning. It maximizes the similarity of positive sample pairs while minimizing the similarity of negative sample pairs, so that the obtained feature representation has stronger discrimination and generalization ability.

\subsection{Motivation example}\label{sec:3-3}

We provide an example to illustrate the limitations of state-of-the-art threat hunting methods based on query graphs and the intuition behind our approach. The example is from DARPA TC Engagement \#3 CADETS. As shown in Fig.~\ref{fig_2}, the provenance graph records the process of the attacker injecting Dragon APT into the server memory through the Nginx backdoor. In the graph, diamonds represent IP addresses, rectangles represent processes, and ellipses represent files. The attacker sends a malformed HTTP request to connect the vulnerable Nginx server to the shellcode server at \textit{78.205.235.65}. Using the shell, the attacker downloads a malicious payload to \textit{/tmp/vUgefal} and escalates privileges with \textit{passwd}. The elevated process then connects to \textit{61.167.39.128}, downloading another malicious payload to \textit{/var/log/devc}. The lower part of the provenance graph records the normal system behavior of mail sending.

\subsubsection{Limitations of existing threat hunting works} 

With the increasing complexity and diversification of network attacks, existing threat hunting methods based on query graphs face increasing limitations, posing challenges for threat detection.

\begin{enumerate}[(1)]
\item 
Laborious query graph generation. Query graphs are generated through the secondary processing of original threat intelligence. Converting natural language text into graphs requires annotators with rich cybersecurity knowledge. The quantity and quality of query graphs are critical for ensuring accurate threat detection, making the creation of high-quality, large-scale query graph libraries a significant challenge. While some researchers \cite{satvat2021extractor,li2022attackg} have attempted to automate this conversion process using NLP techniques, a good NLP model still requires a lot of annotated data. This does not fundamentally solve the labor-intensive problem of query graph generation, and manual annotation and context understanding are still essential.
\item 
Expensive querying process. APT attacks are often concealed within a large number of benign behaviors. For example, the CADETS dataset contains over ten million log entries, yet only about a hundred are related to attacks, accounting for just 0.001\% of the entire log sequence. Query graph-based methods require evaluating potential alignments across the entire provenance graph. In Fig.~\ref{fig_2}, nodes with the same color in both the provenance and query graphs represent corresponding matches. To reduce search complexity, Poirot \cite{milajerdi2019poirot} estimates APT entry points to filter candidate query nodes. However, due to the complexity of the graph structure, this path-based method still requires multiple traversals. DeepHunter \cite{wei2021deephunter} identifies subgraphs around IOCs and uses graph representation learning to compute semantic similarities. While semantic-based method improves subgraph matching efficiency, it still requires exhaustive pairwise comparisons of query and sampled subgraphs for complete determination. ProvG-Searcher \cite{altinisik2023provg} proposes a dynamic programming algorithm for subgraph partitioning, reducing query computational complexity.  However, the large scale and high connectivity of provenance graphs continue to pose significant challenges to graph representation learning.
\item 
Poor interpretability. Threat intelligence contains extensive analyses of attack techniques and strategies by security experts, which help analysts effectively identify and understand network threats. For example, the two threat intelligence instances shown in Fig.~\ref{fig_2} describe the process in which attackers escalate privileges and download malicious payloads. These descriptions not only outline the attack process but also highlight attack techniques and strategies using keywords such as ``elevated'' and ``root access''. In contrast, query graph only provides templates for threat detection, lacking semantic information related to attack techniques. As a result, manual analysis of the identified threat behaviors is still required. These approaches fail to fully leverage threat intelligence for comprehensive threat interpretation.
\end{enumerate}
\subsubsection{Intuition of our approach}

The core idea behind our approach is to take advantage of contrastive learning to align log sequences and threat intelligence describing the same attack pattern into a shared alignment space. By fitting the matched sample pairs and staying away from the mismatched sample pairs, the model can automatically learn the semantic association between intelligence and log, avoiding the high cost of manually designing the query graph. Contrastive learning can automatically discover the latent structure of data and generalize the learned attack patterns to other semantically similar behaviors. As illustrated in Fig.~\ref{fig_2}, the log behavior of THEIA and CADETS datasets are different, but they both represent privilege escalation, malicious server connection, and payload download. CLIProv can use the semantic similarity of intelligence to gather the log sequences with the same attack pattern in a similar semantic space, and help the model understand the key behavior logic corresponding to a specific attack pattern. In the threat query phase, the model can use the attack pattern information in the intelligence to identify the potential attack behavior even if the attack process description is lack of perfect matching. Additionally, CLIProv introduces a subgraph partitioning method based on behavior dependencies and time density, and maps high-dimensional raw data to a low-dimensional semantic space through a semantic search mechanism, reducing the complexity of threat queries. By using threat intelligence as a medium, CLIProv abstracts underlying attack behaviors into more interpretable attack techniques, enhancing the efficiency of threat identification and providing security analysts with a more intuitive and comprehensive defense perspective.

\section{Threat model}
\label{sec:4}

We assume that the provenance data from the host system is complete and has not been maliciously altered or forged by attackers. We also do not consider hardware-level, side-channel, or covert channel attacks, as these behaviors are typically not captured by kernel-level audit systems. CLIProv is a knowledge-based threat hunting method, and we assume that the attacker’s behavior will retain certain attack patterns within the provenance data. Similar attack techniques are reflected in similar attack sequences. For example, in Fig.~\ref{fig_2}, the CADETS dataset shows attackers implanting a backdoor via Nginx, while the THEIA dataset shows a similar process via Firefox. Both follow the core pattern of shell connection $\to$ payload download $\to$ privilege escalation $\to$ process execution. Frequent changes in tactics and techniques pose significant challenges for adversaries. Therefore, we further assume that the fundamental characteristics of attack behavior will largely remain unchanged. However, CLIProv is limited by the scope of the learned threat intelligence, making it unable to detect attacks that significantly deviate from known patterns. This limitation is common in knowledge-based detection systems \cite{altinisik2023provg,milajerdi2019poirot,wei2021deephunter}, which we will discuss further in Section \ref{sec:7}. Despite this, as shown in Section \ref{sec:6_2}, CLIProv demonstrates good generalization for attack patterns with minor variations.

\section{System design and methodology}
\label{sec:5}

\subsection{Overview}
\label{5_1}

\begin{figure*}[!t]
\centering
\includegraphics[width=7in]{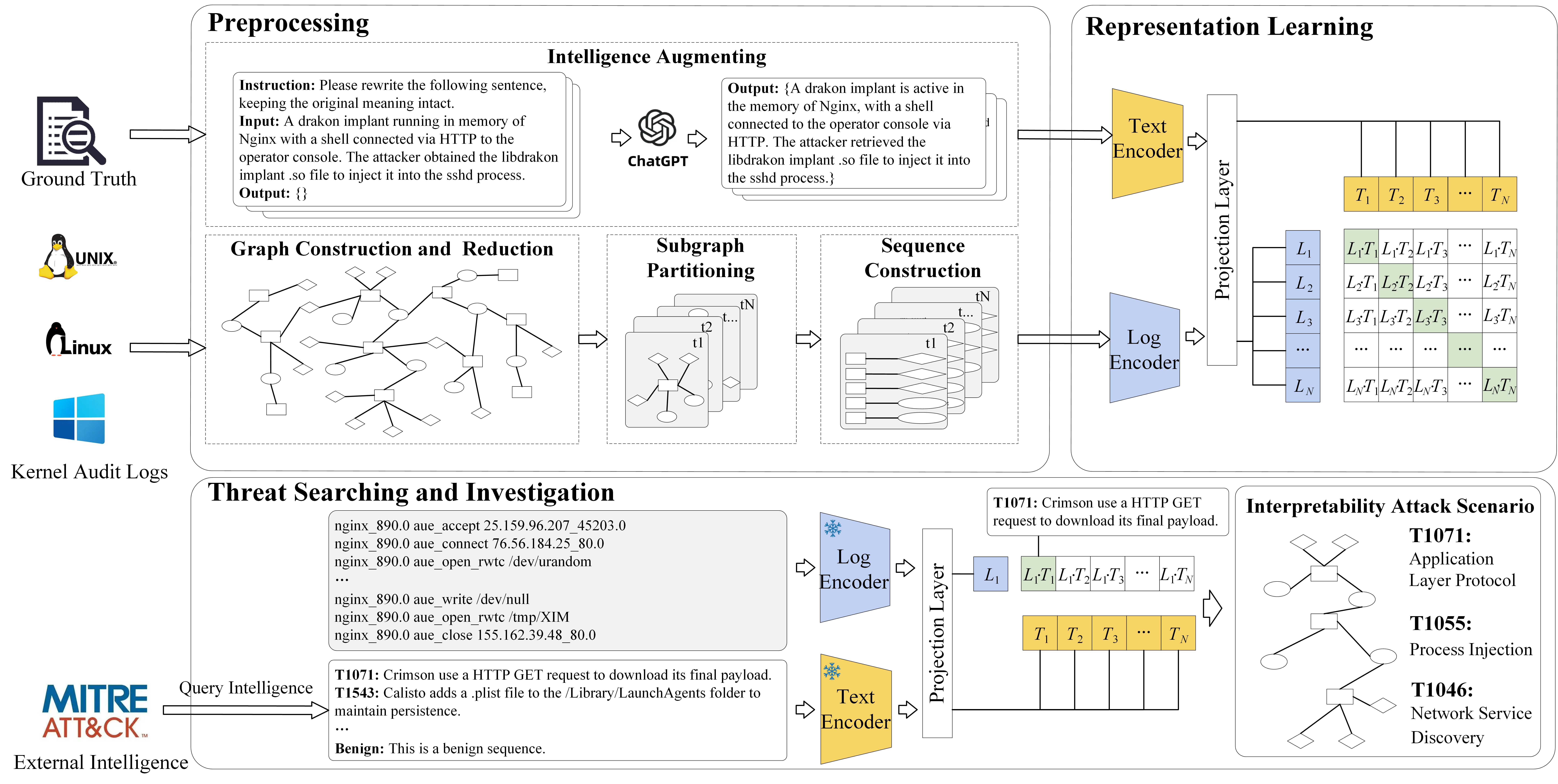}
\caption{Overview of CLIProv’ architecture.}
\label{fig_3}
\end{figure*}

CLIProv is a knowledge-based threat hunting system that employs multimodal learning to achieve semantic alignment between provenance logs and threat intelligence. By leveraging threat intelligence enriched with attack patterns, CLIProv is able to search for potential attack techniques from logs. Fig.~\ref{fig_3} illustrates the architecture of CLIProv, which consists of three main components.

\begin{enumerate}[(1)]
\item 
Preprocessing. CLIProv automatically processes provenance logs and threat intelligence to generate positive and negative sample pairs for representation learning.
\item 
Representation learning. CLIProv employs the pre-trained model Robustly Optimized BERT Approach (RoBERTa) \cite{liu2019roberta} to construct encoders for both threat intelligence text and log sequences, generating vector representations for each. Through a projection layer, data from different modalities are projected into a unified semantic space. Using contrastive learning, the text encoder and log encoder are jointly trained to generate similar vector representations, thereby aligning matching text-log pairs in the semantic space.
\item 
Threat searching and investigation. CLIProv detects threats by querying log sequences in the provenance logs that are semantically similar to threat intelligence. MITRE ATT\&CK \cite{MITRE} is a publicly available threat intelligence repository that documents adversarial tactics, techniques, and common knowledge. Using the trained encoders, the system searches the ATT\&CK database for intelligence that semantically matches the log sequences, thereby identifying attack techniques in the provenance logs. After performing a semantic search across the entire provenance graph, CLIProv designs a temporal causal reasoning method to reconstruct attack scenarios, providing a comprehensive and concise view of the threat landscape.
\end{enumerate}
\subsection{Preprocessing}
\label{5_2}

CLIProv processes provenance logs from various sources, including Unix, Linux, and Windows, to generate log provenance graphs. These graphs are reduced and partitioned based on temporal and spatial features to obtain provenance subgraphs and construct log sequences. Using attack process records from the ground truth, CLIProv generates pairs of threat intelligence and log sequences.

\subsubsection{Graph construction and reduction}
\label{5_2_1}

A provenance graph is a directed graph representing the causal relationships and data flow of events within a system. The provenance graph can be formalized as a tuple ${{G}_{prov}}=(V,E)$, where,

$V$ is the set of vertices representing objects in the system. CLIProv focuses on three types of kernel objects, including processes, files, and network sockets.

$E$ is the set of directed edges representing operations and causal relationships between objects, with the source node denoting the subject initiating the event and the target node representing the object. The edge types indicate event types such as open, read, execute, and connect. Timestamps serve as edge attributes to facilitate graph reduction and partitioning. 

Provenance logs record detailed operations on system objects, often generating numerous duplicate and redundant entries that hinder efficient threat monitoring. To address this, CLIProv reduces the directed disconnected graph by focusing on three key aspects.

\begin{figure}[!t]
\centering
\includegraphics[width=3.5in]{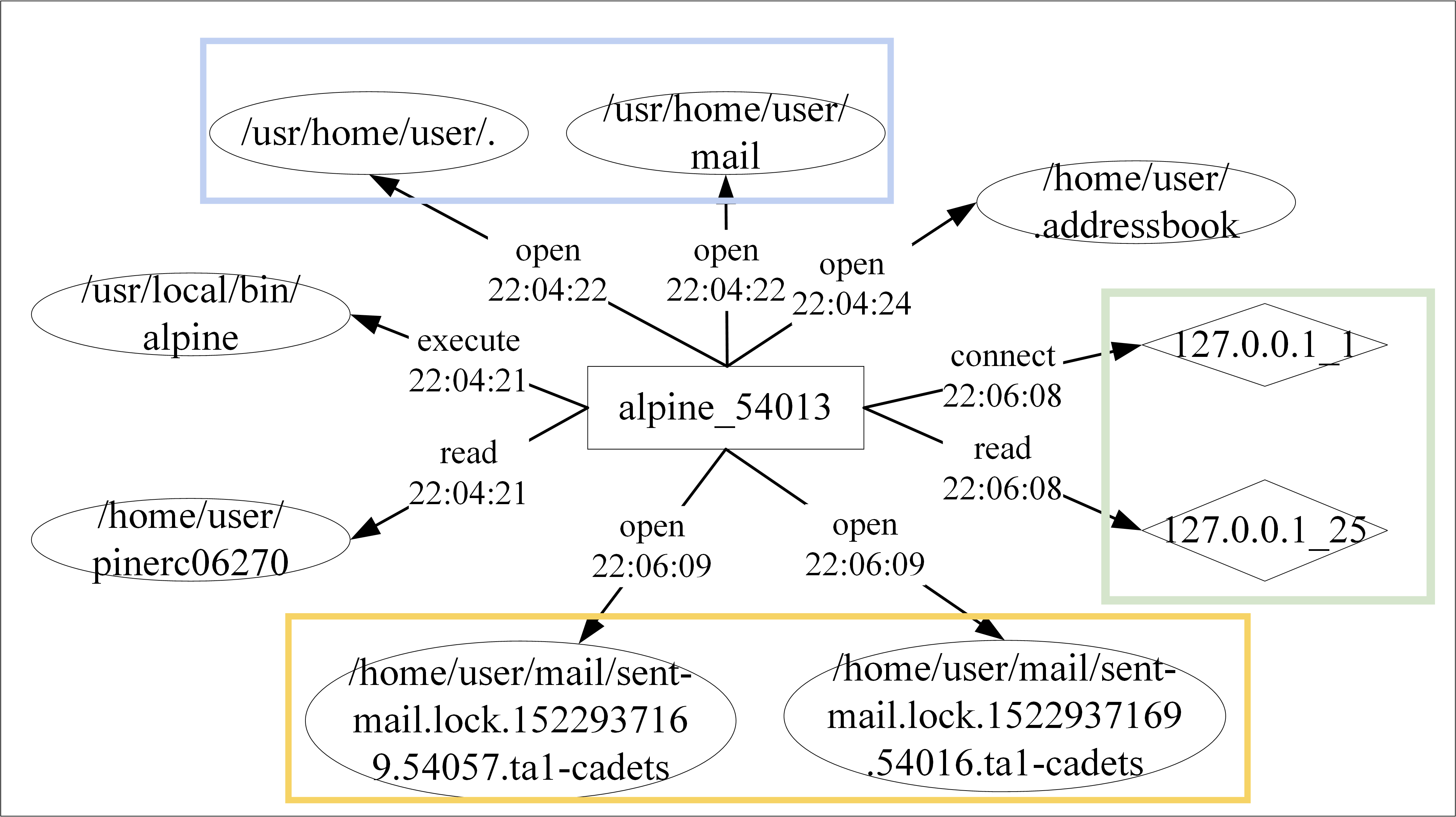}
\caption{Process flow diagram of an alpine email client operation.}
\label{fig_4}
\end{figure}

\begin{enumerate}[(1)]
\item 
Network connections and communications. Within the host system, many processes communicate with other services, such as connection establishment and data transfer, and these operations are usually recorded. In the case of network instability, the system may repeatedly try to reconnect, resulting in a large number of similar entries in the provenance logs. Moreover, a process may access multiple ports in a short period of time during a single communication session. As the example within the green box in Fig.~\ref{fig_4}, the Alpine email client first connects to \textit{127.0.0.1} via \textit{port 1} for an initial network check and subsequently connects to the local mail server via \textit{port 25} to send emails. These two actions happened within one second and before and after each other, so we think that these two actions can be combined into the same action. Both operations occur within one second. CLIProv merges entities and operational behaviors by analyzing the communication from the same process to the same IP address within a fixed time window. If there is access to multiple different ports, multiple ports are connected with \_. If the types of edges are different, the different event types are connected with \_ to form a new edge. The merged timestamp is defined as the earliest timestamp in the set. In Fig.~\ref{fig_4}, the two edges in the green box are merged into a single edge with the event type ``connect\_read'' and a timestamp of ``22:06:08''. 
\item 
File directories cascading operations. In the host system, files need to be accessed step by step, so the cascading operations on file directories often appear in the provenance logs, leading to a lot of redundancy. The blue box in Fig.~\ref{fig_4} records the sequence of Alpine process accesses to the folder \textit{sent}. In this case, \textit{/usr/home/user} is the parent directory of \textit{/usr/home/user/mail}. Accessing the parent directory does not contribute additional value for tracing system operation flows. CLIProv eliminates events associated with cascading directory operations from the provenance graph, retaining only those corresponding to the final directory level.
\item 
Similar file objects. There are often file entities in the host system that have similar functions but are named slightly differently. They usually appear to have the same parent directory with a slight difference in file naming. These are usually temporary files, log files, or cache files generated automatically by the system. These files are created when certain tasks are performed to record state, store intermediate data, or support process operations. Their names tend to follow certain patterns, such as adding a timestamp, process ID, or random string that identifies where the file came from or when it was generated. For example, the locked mail in the user mail folder in the yellow box in Fig.~\ref{fig_4}. To mitigate this redundancy, CLIProv calculates the string similarity of file nodes connected to the same source node and sharing the same parent directory within a fixed time window. String similarity is measured using the Levenshtein distance \cite{levenshtein1966binary}. Nodes with a similarity score exceeding 0.7 are merged into a single entity. 
\end{enumerate}

This approach allows the provenance graph to capture system behaviors comprehensively while preserving a concise structure.

\subsubsection{Subgraph partitioning}
\label{5_2_2}

Threat intelligence generally describes a single attack technique, while the provenance log usually contains the full operation behavior of the system for a long time. To effectively match threat intelligence text, the provenance graph needs to be partitioned into finer-grained subgraphs. A straightforward approach is to partition based on a time window \cite{HanP0MS20}. However, this approach may disrupt the causal relationships between events, leading to fragmented attack scenarios. Determining an appropriate time span is also challenging, as a span that is too short may miss critical steps, while one that is too long may introduce irrelevant noise and reduce the accuracy of threat identification.

According to the principle of temporal locality in operating systems and the assumption of behavioral consistency, the events involved in a system behavior are logically highly correlated and exhibit temporal density. After process A creates process B, B's access to file C usually occurs within a short period of time. In contrast, there are logical gaps and time intervals between different behaviors, with prolonged pauses often indicating task switching or condition waiting. During actual attacks, APT attackers often adopt multi-step and low-frequency strategies to enhance the stealth and persistence of their actions. They deliberately extend the time intervals between attack steps to evade security detection. Therefore, CLIProv defines system behavior subgraphs as the smallest unit for training. A system behavior subgraph represents a series of operations performed by a specific process on different objects within a continuous time period. 

For process $p$ and its set of interacting objects, its system behavior subgraph $G(p)$ is defined as a sequence of events that satisfies the following conditions:

\begin{equation}
\label{eq_10}
G(p)=(V(p),E(p),T(p))
\end{equation}

Where $V(p)$ represents the set of nodes, including $p$ and the set of its accessed objects $O=\{{{o}_{1}},{{o}_{2}},... ,{{o}_{n}}\}$. $E(p)$ represents the set of interaction events. $(p)$ represents the set of timestamp, satisfying the time continuity and density. For any two consecutive events ${e}_{i}, {e}_{i+1} \in E(p)$, the following formula holds:

\begin{equation}
\label{eq_11}
\Delta t=t({{e}_{i+1}})-t({{e}_{i}})\le {{\theta }_{\max }}
\end{equation}

${\theta }_{\max }$ is the time threshold used to define the behavior boundary, which can be dynamically adjusted based on the specific scenario. 

First, in order to maintain the integrity of system behavior, a Depth-First Search (DFS) \cite{tarjan1972depth} is used to partition the graph. Starting from an unvisited node in the provenance graph, all subsequent nodes are recursively visited along the direction of the edges until all reachable nodes are visited. The visited nodes and their connecting edges form a dependency subgraph. Then, based on the temporal density characteristics of the events, the time intervals between events are analyzed, and the dependency subgraph is further divided into different system behavior subgraphs. By traversing all nodes in the provenance graph, the final set of system behavior subgraphs is generated. The subgraph partitioning algorithm is detailed in Algorithm \ref{alg1}.

\begin{algorithm}[H]
\caption{Subgraph partitioning algorithm.}\label{alg:alg1}
\begin{algorithmic}
\STATE 
\STATE visited $\gets$ set()
\STATE \textbf{def} {\textsc{DFS}}(${G}_{prov}$, node, visited, $V'$)
\STATE \hspace{0.5cm} \textbf{add} node \textbf{to} visited
\STATE \hspace{0.5cm} \textbf{add} node \textbf{to} $V'$
\STATE \hspace{0.5cm} \textbf{for} neighbor \textbf{in} ${G}_{prov}$.successors(node)
\STATE \hspace{1cm} \textbf{if} neighbor \textbf{not in} visited
\STATE \hspace{1.5cm} \textsc{DFS}(${G}_{prov}$, neighbor, visited, $V'$)
\STATE \hspace{1cm}  \textbf{else}
\STATE \hspace{1.5cm} \textbf{add} neighbor \textbf{to} $V'$
\STATE \textbf{for} node \textbf{in} $V$
\STATE \hspace{0.5cm} \textbf{if} node \textbf{not in} visited
\STATE \hspace{1cm}  $V'$ $\gets$ set()
\STATE \hspace{1cm}  {\textsc{DFS}}(${G}_{prov}$, node, visited, $V'$)
\STATE \hspace{1cm}  $E' \gets \left\{ (u, v) \mid (u, v) \in E \ \text{and} \ u \in V' \ \text{and} \ v \in V' \right\}$ 
\STATE \hspace{1cm}  \textbf{sort} $E'$ \textbf{by} timestamp
\STATE \hspace{1cm}  start\_index $\gets$ 0
\STATE \hspace{1cm}  \textbf{for} $i = 1,...,len(E')$ 
\STATE \hspace{1.5cm}  $\Delta \gets E'[i] - E'[i-1]$
\STATE \hspace{1.5cm}  \textbf{if} $\Delta \geq threshold$
\STATE \hspace{2cm}  $E'' \gets E'[ start\_index:i]$
\STATE \hspace{2cm}  \textbf{add} ${G}_{prov}(E'')$ \textbf{to} sub\_G
\STATE \hspace{2cm}  start\_index $\gets i$
\STATE \textbf{return}  sub\_G
\end{algorithmic}
\label{alg1}
\end{algorithm}

The time complexity of this algorithm is $O(V+E\log E)$, making it significantly more efficient than typical quadratic or polynomial time complexities. This efficiency ensures its suitability for partitioning large-scale provenance graphs.

\subsubsection{Sequence construction}

We convert the system behavior subgraphs into log sequences denoted as $L=\{{{l}_{1}},{{l}_{2}},...,{{l}_{N}}\}$. The log sequence ${{l}_{i}}=[({v_{src}}_{(1,i)},{{e}_{(1,i)}},$ ${v_{dest}}_{(1,i)}),({v_{src}}_{(2,i)},{{e}_{(2,i)}},{v_{dest}}_{(2,i)}),...,$ $({v_{src}}_{({{\mathbb{N}}_{i}},i)},{{e}_{({{\mathbb{N}}_{i}},i)}},{v_{dest}}_{({{\mathbb{N}}_{i}},i)})]$, where ${{\mathbb{N}}_{i}}$ is the number of behavior triples in the $i$-th sequence, ${v_{src}}_{(1,i)}$ is the source node name of first behavior, ${e}_{(1,i)}$ is the edge type, and ${v_{dest}}_{(1,i)}$ is the destination node name.

\subsubsection{Intelligence augmenting}

We manually divided the threat intelligence into the ground truth according to the generated attack log sequence. Take the attack scenario in Fig.~\ref{fig_1} as an example, the above subgraph partitioning algorithm divides the attack scenario into two subgraphs, and the corresponding threat intelligence is also divided into two parts. The threat intelligence corresponding to the log sequence about the nginx process is ``\textit{Began attack with CADETS FreeBSD by exploiting Nginx. The attack attempt succeeded and resulted in loaderDrakon connected to an operator console shell (78.205.235.65). The attacker downloaded a file to be elevated as a new process running as root.}'' 

Given the limited number of attack sequence samples, CLIProv expand the number of sample pairs by enhances threat intelligence. In recent years, large language models have demonstrated broad application potential in text data augmentation due to their powerful few-shot learning capability \cite{HU2024103999,10489969}. CLIProv leverages a prompt template and GPT-3 \cite{NEURIPS2020_1457c0d6} to augment threat intelligence. The prompt template is defined as ``\textit{Please rewrite the following sentence, keeping the original meaning intact.}'' We input the initial, unaltered threat intelligence into the GPT model, which generates augmented threat intelligence data. This method expands threat intelligence to obtain more sample pairs, helping the model learn more comprehensive and generalizable features. 

For benign log sequences, we use ``\textit{This is a benign sequence.}'' as their corresponding textual description. The threat intelligence sample set is denoted as $T=\{{{t}_{1}},{{t}_{2}},...,{{t}_{M}}\}$, where ${{t}_{j}}$ is the $j$-th intelligence text.

\subsection{Log sequence representation learning}

CLIProv employs a dual-encoder architecture to separately encode provenance logs and threat intelligence samples. A projection layer is designed to map the features from different modalities into a shared feature space, where contrastive learning is used to align and fit the features across modalities.

\subsubsection{Encoder} 

As shown in Fig.~\ref{fig_3}, CLIProv consists of a text encoder and a log encoder. We treat log sequences as sentences composed of system entities and events, utilizing a pre-trained RoBERTa model as the log encoder. Similarly, the text encoder also employs RoBERTa. We pair samples from L and T to form a sample set for multimodal training. Corresponding intelligence-log pairs are considered positive samples, while others are considered negative samples. For each log sequence $l\in L$, we encode it with the log encoder ${{f}_{\theta }}(l)$ parameterized by $\theta $ , and get the “[CLS]” feature as the vector representation of the log sequence $v={{f}_{\theta }}(l)\in {{\mathbb{R}}^{d}}$. For each intelligence text $t\in T$, we encode it with the text encoder ${{f}_{\phi }}(t)$ parameterized by $\phi $ , and get the “[CLS]” feature as the vector representation of the intelligence text $u={{f}_{\phi }}(t)\in {{\mathbb{R}}^{d}}$.

\subsubsection{Projection network} 

To achieve the alignment of log sequences and threat intelligence in the shared semantic space, we define a two-layer neural network as a projection network. To mitigate gradient vanishing issues and improve network convergence speed, we utilized residual connections to project data from both modalities into the same vector space. Given an input vector $x\in {{\mathbb{R}}^{d}}$, the output of the first layer network is ${{h}_{1}}=\sigma ({{W}_{1}}x+{{b}_{1}})$, where $\sigma $ is the activation function (e.g., ReLU), ${{W}_{1}}\in {{\mathbb{R}}^{{{d}_{out}}\times d}}$ is the weight matrix of the first layer, ${{b}_{1}}\in {{\mathbb{R}}^{{{d}_{out}}}}$ is the bias vector. The output of the residual connection is $g(x)={{h}_{1}}+\sigma ({{W}_{2}}{{h}_{1}}+{{b}_{2}})$. After the projection layer, we obtain vector representations for the log sequences and threat intelligence, respectively as $g({{f}_{\theta }}(l))$and $g({{f}_{\phi }}(t))$.

\subsubsection{Training} 

For $i$-th log ${{l}_{i}}$ and $j$-th intelligence ${{t}_{j}}$, we normalize their feature vectors using ${{v}_{i}}=\frac{g({{f}_{\theta }}({{l}_{i}}))}{\left\| g({{f}_{\theta }}({{l}_{i}})) \right\|}$ and ${{u}_{j}}=\frac{g({{f}_{\phi }}({{t}_{j}}))}{\left\| g({{f}_{\phi }}({{t}_{j}})) \right\|}$, and their similarity is calculated as ${{v}_{i}}^{T}{{u}_{j}}$. A bidirectional supervised contrastive objective is considered to train the model, that uses InfoNCE \cite{oord2018representation} as the loss function to enhance model robustness. Here, the log-to-text loss function is defined as:

\begin{equation}
\label{eq_3}
Los{{s}_{l2t}}=-\sum\limits_{i=1}^{N}{\log \frac{\exp ({{v}_{i}}^{T}{{u}_{i}}/\tau )}{\sum\nolimits_{j=1}^{N}{\exp ({{v}_{i}}^{T}{{u}_{j}}/\tau )}}}
\end{equation}

and the text-to-log loss function is defined as:

\begin{equation}
\label{eq_4}
Los{{s}_{t2l}}=-\sum\limits_{i=1}^{N}{\log \frac{\exp ({{u}_{i}}^{T}{{v}_{i}}/\tau )}{\sum\nolimits_{j=1}^{N}{\exp ({{u}_{i}}^{T}{{v}_{j}}/\tau )}}}
\end{equation}

where $\tau$ is the learnable temperature hyper-parameter \cite{wu2018unsupervised} to scale the logits, which is initialized to 0.07. The final loss is the average of the two:

\begin{equation}
\label{eq_5}
Loss=\frac{Los{{s}_{l2t}}+Los{{s}_{t2l}}}{2}
\end{equation}

CLIProv is further pre-trained from the original pre-trained model, allowing us to utilize the log encoder and text encoder to obtain vector representations of log sequences and threat intelligence.

\subsection{Threat searching and investigation}

The goal of threat search and investigation is to discover and reconstruct the entire attack process within a complete provenance scenario. First, the provenance graph is preprocessed and partitioned into system behavior subgraphs, which are then used to obtain log sequences. Next, the log encoder generates vector representations of these sequences. By searching the most semantically similar intelligence texts from the query database, CLIProv identifies all suspicious log sequences, thereby reconstructing the entire attack scenario.

CLIProv constructs a query database $Q=\{{{Q}_{1}},{{Q}_{2}},...,$ ${Q}_{K}\}$, using the MITRE ATT\&CK threat intelligence repository. For the $k$-th query ${{Q}_{k}}=({{q}_{k}},{{y}_{k}})$, ${{q}_{k}}$ is the text description, ${{y}_{k}}$ is the intelligence label. If ${{q}_{k}}$ is a threat intelligence, ${{y}_{k}}$ is its associated TTP. For example, ${{Q}_{k}}=$(``\textit{Crimson can use a HTTP GET request to download its final payload.}'', “\textit{T1071}”). If ${{q}_{k}}$ is ``\textit{This is a benign sequence.}'', ${{y}_{k}}$ is ``\textit{benign}''. The query text is fed into the pre-trained text encoder to obtain the query vector ${{f}_{\phi }}({q}_{k})$. The log sequence $l$ is encoded with ${{f}_{\theta }}(l)$. In the alignment space, CLIProv selects the query vector with the smallest semantic distance to $l$ and obtains its associated intelligence label.

\begin{equation}
\label{eq_6}
\hat{y} = y_{\arg \min_k \left( g\left( f_{\theta}(l) \right)^T g\left( f_{\phi}(q_k) \right) \right)}
\end{equation}

CLIProv identifies all log sequences labeled with TTPs as attacks. By searching the whole provenance graph, all attack sequences can be obtained, which can be represented as a set of attack sequences $L_{attack}=\{{l}_{1},{l}_{2},...,{l}_{n}\}$. Each log sequence ${l}_{i}$ represents a system behavior subgraph $G(i)$. To connect these isolated subgraphs, CLIProv considers the provenance relationships between attack behaviors and proposes an efficient temporal causal reasoning method. It introduces a virtual node $v_i'$ for each subgraph, and connects $v_i'$ with all other nodes $v\in V(i)$ in the subgraph. The generated virtual subgraph is represented as follows.

\begin{equation}
\label{eq_7}
G(i)' = (V(i) \cup \{v_i'\}, E(i) \cup \{(v_i', v) \mid v \in V(i)\}, t_i)
\end{equation}

where $t_i$ represents the timestamp of the subgraph, taking the earliest occurrence time of the edges in the subgraph to indicate the time of the virtual node.

CLIProv uses virtual nodes to represent behavior subgraphs and reconstructs attack paths based on the causal relationships between virtual nodes. Specifically, CLIProv sorts the $v_i'$ by $t_i$. For each $v_i'$, reason about the causal relationship between it and the virtual node at the previous time step. If there is no direct correlation between adjacent virtual nodes, CLIProv traverses forward until it finds a virtual node that has a causal relationship with $v_i'$. The causal relationship is the shortest attack path between virtual nodes, which is computed by a time-dependent Dijkstra algorithm \cite{ding2008finding}. It ensures they are connected in the order of the attack occurrence. The identified attack paths are as follows.

\begin{equation}
\label{eq_8}
{{P}_{i\_prev,i}}=\arg {{\min }_{{{v}_{i\_prev}}\in {{V(i\_{prev})}},{{v}_{i}}\in {{V(i)}}}} \text{distance} (v_{i\_prev}',v_{i}')\
\end{equation}

where $v_{i\_prev}'$ denotes the closest virtual node with predecessor relationship to $v_{i}'$, $t_{i\_prev}<t_i$.

Through continuous iteration, CLIProv connects all isolated subgraphs, forming a complete attack graph ${{G}_{attack}}$. The reconstruction process is as followed.

\begin{equation}
\label{eq_9}
{{G}_{attack}}=\bigcup\limits_{i=2}^{n}{(({G(i\_prev)'}\backslash \{v_{i\_prev}'\})\cup ({G(i)'}\backslash \{v_{i}'\})\cup {{P}_{i\_prev,i}})}
\end{equation}

where $\backslash$ means to remove the virtual node from the graph.

Through temporal causal reasoning, CLIProv can reconstruct isolated behavior sequences into a complete attack scenario. The TTPs labels within the subgraphs provide high-level tactical and technical knowledge, helping security analysts quickly identify attack patterns and targets, thereby enabling more effective defense.

\section{Evaluation}
\label{sec:6}

We developed a prototype of CLIProv and evaluated it using four public datasets. This section first describes our experimental setup, including the development framework, runtime environment, and datasets. Then, We evaluate the effectiveness and generalizability of CLIProv, comparing it with four other threat investigation methods. Finally, we evaluate the impact of various parameters, runtime overhead, and the reliability of the generated interpretable attack scenarios.

\subsection{Experiment setup}

We implemented a CLIProv prototype in Python 3.8. We use NetworkX \cite{ding2008finding} to implement provenance graph processing and PyTorch \cite{paszke2019pytorch} to implement the encoder model. All experiments are conducted on a Ubuntu 20.04 machine equipped with a GeForce RTX A6000 GPU, 80 3.10GHz CPUs, and 1TB of main memory.

\subsubsection{Datasets}

We evaluate our approach on CADETS, THEIA, ATLAS, and CICAPT-IIoT datasets, and use the ATT\&CK database as the query intelligence repository. We reduced the size of the provenance graphs for the log datasets and selected specific attack scenarios for training CLIProv, while other attack scenarios were used to test the model's effectiveness. Table~\ref{tab_1} summarizes the statistics of the datasets. 

\begin{table*}[htbp]
  \centering
  \caption{Statistics of the log datasets}
  \label{tab_1}
    \begin{tabular}{p{6em}p{2.5em}p{2.5em}p{2.5em}p{4.5em}p{4.5em}p{2em}p{2em}p{4.5em}p{5em}}
    \toprule
    \textbf{Datasets} & \textbf{Original Nodes} & \textbf{Original Edges} & \textbf{Size} & \textbf{Preprocessed Nodes} & \textbf{Preprocessed Edges} & \textbf{Train Graph} & \textbf{Test Graph} & \textbf{Threat Intelligence} & \textbf{Query Intelligence} \\
    \midrule
    CADETS      & 454.1K & 1.1M  & 11.1GB & 236.7K & 673.4K & 2   & 2   & 3    & 10358 \\
    \cmidrule{1-9}
    THEIA       & 426.8K & 1.6M  & 43.2GB & 354.9K & 743.4K & 2   & 1   & 3    &       \\
    \cmidrule{1-9}
    ATLAS       & 200.9K & 2.5M  & 4.1GB  & 76.5K  & 1.4M   & 10  & 6   & 6    &       \\
    \cmidrule{1-9}
    CICAPT-IIoT & 53.3K  & 143.5K& 28.5MB & 27.9K  & 92.4K  & 0   & 1   & 0    &       \\
    \bottomrule
    \end{tabular}
\end{table*}

CADETS \cite{transparent} is from the 3rd engagement of the DARPA TC program and includes two weeks of system logs documenting red team versus blue team scenarios. The simulated attack activities involve penetrating a FreeBSD host through an Nginx backdoor and deploying malicious payloads. Attackers conducted four attacks, each lasting no more than one hour. DARPA provided ground truth for these attacks, allowing us to label the subgraphs associated with the attacks and obtain threat intelligence. We also collected attack scenario descriptions from KAIROS \cite{KAIROS}. To evaluate CLIProv's generalization ability across attack scenarios, we selected two scenarios from CADETS as the training set and two as the testing set.

THEIA \cite{transparent} is also from the 3rd engagement of the DARPA TC program and documents three attacks on Ubuntu hosts. These attacks utilized three different methods: Firefox backdoor, browser extension, and phishing. The attack labeling and threat intelligence collection processes are consistent with those used in the CADETS dataset. We selected two scenarios from THEIA as the training set and one as the testing set.

ATLAS \cite{alsaheel2021atlas} is from ATLAS and contains publicly available simulated attacks on Windows hosts. These attacks were reproduced based on APT reports and include sixteen incidents, which exploit six vulnerabilities: CVE-2015-5122, CVE-2015-3105, CVE-2015-5119, CVE-2017-11882, CVE-2017-0199, and CVE-2018-8174. We used the disclosed attack nodes to label the subgraphs related to the attacks. The corresponding threat intelligence for the attack subgraphs was generated manually based on the referenced APT reports. We selected attack scenarios related to CVE-2015-5122 and CVE-2015-5119 in the ATLAS dataset as the testing set, with the rest as the training set.

CICAPT-IIoT \cite{ghiasvand2024cicapt} is captured by the University of New Brunswick in an industrial IoT environment, focusing on the detection and analysis of APT attacks. The dataset was collected over three days. It simulates APT29 techniques executed via Kali VM1 and MITRE Caldera, encompassing over 20 different attack techniques and 8 attack tactics. Since the dataset does not provide ground truth for the attack processes, it is used solely to test the generalization ability of CLIProv.

ATT\&CK \cite{MITRE} is a TTP dataset that has been annotated by multiple cybersecurity experts, containing descriptions and examples of tactics and techniques. It includes 14 tactics, 177 techniques, and 10358 threat intelligence.

\subsubsection{Parameter setup}

To achieve finer-grained partitioning of log sequences in the provenance graph, we employed the subgraph partitioning algorithm described in Section \ref{5_2_2} to transform the provenance graphs into system behavior subgraphs. The behavioral boundary threshold is set as ${\theta }_{\max }=20min$. To address the imbalance between positive and negative samples in the provenance logs, we used GPT-3 to augment threat intelligence texts. The number of augmented texts is ${{n}_{aug}}=3$. The detailed statistical information of the final dataset is presented in Table~\ref{tab_2}.

For the CLIProv model, we selected the pre-trained RoBERTa model to serve as both the log encoder and text encoder. RoBERTa, known for its improved stability and generalization compared to BERT, was used with embedding dimensions of 768 for both logs and texts. The projection layer consisted of a two-layer neural network with 128 hidden units in each layer. During training, we used a batch size of 64 and performed 100 epochs. To select the optimal hyperparameters, we used a grid search strategy, with an initial learning rate of 1e-5 and a dropout rate of 0.5.

\begin{table*}[htbp]
  \centering
  \caption{Statistics of the final datasets}
  \label{tab_2}
    \begin{tabular}{p{6em}p{6em}p{6em}p{6em}p{6em}p{5em}}
    \toprule
    \textbf{Datasets} & \textbf{Attack Graph for Training} & \textbf{Benign Graph for Training} & \textbf{Attack Graph for Testing} & \textbf{Benign Graph for Testing} & \textbf{Threat Intelligence} \\
    \midrule
    CADETS      & 5      & 24,691  & 4       & 6,043   & 28   \\
    \midrule
    THEIA       & 7      & 8,979   & 3       & 3,824   & 44   \\
    \midrule
    ATLAS       & 37     & 6,001   & 21      & 3,907   & 148  \\
    \midrule
    CICAPT-IIoT & 0      & 0       & 58      & 6,222   & 0    \\
    \bottomrule
    \end{tabular}
\end{table*}

\subsection{Effectiveness}
\label{sec:6_2}

We utilized three training sets: CADETS, THEIA, and ATLAS, along with their respective threat intelligence, to train CLIProv for learning vector representations of log sequences and threat intelligence. The model was then evaluated on the test set. Table~\ref{tab_3} presents the investigation results for each test set under both node-level and graph-level detection. In node-level detection, we label the nodes within the identified attack subgraphs as attack and the nodes within benign subgraphs as benign, then compare the identified attack nodes with the ground truth. In graph-level detection, we compare the identified attack subgraphs with the ground truth attack subgraphs. Table~\ref{tab_11} presents the identified TTP results in different test scenarios. The CICAPT-IIoT dataset contains a large number of subgraphs, and only the top 5 most frequently identified techniques are displayed.

\begin{table*}[htbp]
  \centering
  \caption{The detection results of attacks in different datasets}
  \label{tab_3}
    \begin{tabular}{p{6em}p{11em}p{10em}p{11em}p{10em}}
    \toprule
    \textbf{Datasets} & \textbf{Node-level Precision (\%)} & \textbf{Node-level Recall (\%)} & \textbf{Graph-level Precision (\%)} & \textbf{Graph-level Recall (\%)} \\
    \midrule
    CADETS       & 76.67     & 100       & 100       & 100      \\
    \midrule
    THEIA        & 82.14     & 100       & 100       & 100      \\
    \midrule
    ATLAS        & 62.09     & 100       & 100       & 100      \\
    \midrule
    CICAPT-IIoT  & 52.38     & 84.36     & 60.96     & 86.21    \\
    \bottomrule
    \end{tabular}
\end{table*}

\begin{table}[htbp]
\centering
\caption{TTP results identified in the test scenario}
\label{tab_11}
\begin{tabular}{p{6em}p{2em}p{2em}p{2em}p{2em}p{2em}}
\toprule
\textbf{Datasets} & \textbf{G1} & \textbf{G1}& \textbf{G1}& \textbf{G1}& \textbf{G1}\\
    \midrule
    CADETS\_1       & T1071     & NA       & NA        & NA    & NA    \\
    \midrule
    CADETS\_2       & T1071     & T1055       & T1046        & NA    & NA    \\
    \midrule
    THEIA       & T1189    & T1546       & T1046        & NA    & NA    \\
    \midrule
    ATLAS\_M1h1       & T1204     & T1105       & T1083     & T1083    & NA    \\
    \midrule
    ATLAS\_M1h2       & T1071     & T1083       & NA        & NA    & NA    \\
    \midrule
    ATLAS\_M2h1       & T1569     & T1059       & T1083        & NA    & NA    \\
    \midrule
    ATLAS\_M2h2       & T1569     & T1059       & T1083        & NA    & NA    \\
    \midrule
    ATLAS\_S1      & T1071     & T1041       & T1041       & T1105   & T1083    \\
    \midrule
    ATLAS\_S2      & T1204    & T1569       & T1083         & T1571    & NA    \\
    \midrule
    CICAPT-IIoT      & T1546     & T1055    & T1083        & T1005    & T1041    \\
\bottomrule
\end{tabular}
\end{table}

\begin{enumerate}[(1)]
\item 
Generalization across attack scenarios: Experiments on CADETS and THEIA datasets show that CLIProv has strong generalization ability in cross-scenario attack detection. The model is able to identify potential attacks even if they don't appear in the training data. For example, in the browser extension attack scenario within the THEIA dataset, CLIProv correctly identified the technique \textit{Drive-by Compromise (T1189)}. Although it was unable to precisely identify the corresponding technique \textit{Browser Extension (T1176)}, it successfully identified anomalous behavior by associating the attack behavior with keywords such as ``websites'' and ``browsers''.  This indicates that CLIProv can capture highly correlated patterns with attack behavior in the feature space, thus maintaining a certain detection capability in unknown scenarios.
\item 
Generalization in unknown vulnerabilities: Experiments on ATLAS dataset show that CLIProv has strong generalization ability in cross-vulnerability attack detection. Even when the training set does not contain a specific vulnerability, such as CVE-2015-5122 in the ATLAS M1h1 scenario, CLIProv is able to successfully detect attacks by identifying exploitation-related behavior patterns, including \textit{Malicious Link (T1204)}, \textit{Payload Download (T1105)}, and \textit{File Discovery (T1083)}. This result shows that CLIProv not only relies on known vulnerability features, but also can learn and generalize potential attack patterns in the behavior sequence, so as to maintain the detection ability in unknown vulnerabilities.

\item 
Generalization across datasets: Experimental results on the CICAPT-IIoT dataset show that CLIProv is able to identify most attack behaviors even if the model is not pre-trained on the target dataset. By using contrastive learning to align threat intelligence with provenance logs, CLIProv learns the correlation between attack patterns and system activities. At the same time, the model also learns a large number of benign behaviors in the training process, so that it still has strong discrimination ability when dealing with unknown attacks. However, due to the limitation of knowledge, CLIProv may suffer from false negative when facing techniques with large differences from known attack patterns. For example, the \textit{Wi-Fi Discovery (T1016)} involves network configuration information collection behavior, which accesses the key configuration files \textit{/etc/resolu.conf} and \textit{/etc/hosts} by calling \textit{cmd.exe} to execute the command \textit{ipconfig}. These operations are also common in normal network operation and maintenance scenarios. Due to the lack of similar attack samples, the model cannot distinguish the differences between them and normal network operation and maintenance, resulting in false positives. However, CLIProv still achieved a recall of 86.21\%. While these results indicate some room for improvement, the model's ability to maintain a relatively high recall against unseen attack techniques highlights its generalization ability.
\end{enumerate}
\subsection{Comparison study}
\label{6_3}

In this section, we compare and evaluate CLIProv with other threat detection methods on CADETS and THEIA datasets.

\subsubsection{Query graph-based threat hunting methods}
\label{6_3_1}

As an external intelligence-based threat hunting method, we specifically focus on query graph-based approaches, which utilize threat intelligence as templates for threat queries. We select POIROT \cite{milajerdi2019poirot} and ProvG-Searcher \cite{altinisik2023provg} for comparative evaluation. POIROT is the first method to propose query graphs. By performing graph matching in the provenance graph, POIROT identifies attack subgraphs that are isomorphic to the query graph, which is representative of query graph-based detection methods. ProvG-Searcher is currently the most advanced query graph-based method. Compared to POIROT, ProvG-Searcher introduces GNN for learning representations of the provenance graph and uses subgraph partitioning and semantic matching to identify attack subgraphs. Notably, ProvG-Searcher shares a similar architecture with CLIProv. We evaluate all three methods on the CADETS and THEIA datasets, as the query graphs used in these methods are derived from DARPA's ground truth. Table \ref{tab_5} shows the accuracy and running efficiency of these three methods in graph-level detection.

\begin{table*}[htbp]
  \centering
  \caption{Comparison experiments with query graph-based methods}
  \label{tab_5}
    \begin{tabular}{p{6em}p{4em}p{5em}p{4em}p{5.5em}p{4em}p{7em}p{5em}}
    \toprule
    \multirow{2}{*}{\textbf{Method}} & \multirow{2}{*}{\textbf{Datasets}} & \textbf{Graph-level FPR} & \textbf{Alignment Score} & \textbf{Preprocessing Time} & \textbf{Training Time} & \textbf{Threat searching Time} & \textbf{Total Time} \\

    \midrule
    \multirow{2}{*}{POIROT} 
    & CADETS    & 0    & \textbf{0.89}   & NA      & NA       & 5.45h   & 24.88h \\
    \cmidrule{2-7} 
    & THEIA      & 0    & 0.71   & NA      & NA       & 19.43h  &     \\
    \midrule
    \multirow{2}{*}{ProvG-Searcher}
    & CADETS     & 0.43 & 0.79   & 11 min  & 10.83h    & 29s     & 11.20h \\
    \cmidrule{2-5} \cmidrule{7-7}
    & THEIA      & 0.25 & 0.79   & 11 min  &     & 18s     &      \\
    \midrule
    \multirow{2}{*}{CLIProv}  
    & CADETS     & \textbf{0}    & 0.77   & 11 min  & \textbf{8.72h}    & \textbf{23s}     & \textbf{9.15h}  \\
    \cmidrule{2-5} \cmidrule{7-7}
    & THEIA     & \textbf{0 }   & \textbf{0.84}   & 15 min  &       & \textbf{8s}      &      \\
    \bottomrule
    \end{tabular}
\end{table*}

\begin{enumerate}[(1)]
\item 
Accuracy. In graph-level detection, Poirot and CLIProv have no false positives, demonstrating their high detection accuracy. In contrast, ProvG-Searcher enhances the matching ability by simplifying the entity types in the query graph, but loses highly discriminative features such as process names, filenames, and IP addresses. Although this simplification can improve the matching efficiency of the algorithm, it is easy to lead to false matching. ProvG-Searcher produced three false matches in the CADETS dataset and one false match in the THEIA dataset. To evaluate the quality of the generated attack scenarios, we use the alignment scoring method from POIROT to calculate the alignment scores between the attack scenarios from all three methods and the ground truth graphs. As shown in Table~\ref{tab_5}, even without using precise query templates, CLIProv's alignment scores are comparable to those of the other two methods.

\item 
Efficiency. POIROT employs an exact graph matching algorithm, ensuring high accuracy and low false positives. However, it searches and matches on the whole provenance graph, the time complexity of subgraph matching increases exponentially with the increase of the number of nodes and edges. For example, THEIA dataset contains 426.80K nodes with 1.60M edges, which makes the query time of Poirot reach 19.43 hours. ProvG-Searcher reduces the need for exhaustive computation by embedding subgraphs into a vector space, reducing the matching process to a comparison between the query and the precomputed embedded subgraphs. However, due to the complexity of the provenance graph, GNN-based graph representation learning is more time-consuming than sequence-based models, taking 10.83 hours to train the graph representation model. In comparison, CLIProv takes only 8.72 hours to pre-train, and the detection time is 8 seconds faster than ProvG-Searcher on average. Moreover, both POIROT and ProvG-Searcher use seven specific query graphs for threat search. The query time will further increase when there are multiple query graphs in the system.
\item 
Interpretability. As shown in the upper part of Fig.~\ref{fig_2}, both POIROT and ProvG-Searcher manually extract query graphs from threat intelligence and perform matching within the provenance logs. The attack scenarios generated by these methods only represent the attack execution process and lack inherent interpretability, requiring further analysis by security experts. In contrast, CLIProv can identify potential attack techniques and provide valuable references for threat hunting.
\end{enumerate}
\subsubsection{Anomaly-based detection methods}
\label{6_3_2}

In order to highlight the advantages of the knowledge-driven approach, we compare our approach with two anomaly-based threat detection methods, Threatrace \cite{wang2022threatrace} and KAIROS \cite{KAIROS}. Threatrace and KAIROS model normal nodes within the provenance graph to detect anomalies. Threatrace considers the neighbor nodes around the attack node to also help the attack process, and therefore marks the 2-hop neighbor nodes of the attack node as also anomaly. For a fair comparison, we evaluated the above three methods on the datasets of 2-hop labeled and original labeled, and evaluated their precision, recall, and accuracy at the node level. The comparative results are shown in Table~\ref{tab_6}.

\begin{table*}[htbp]
  \centering
  \caption{Comparison experiments with anomaly-based methods}
  \label{tab_6}
    \begin{tabular}{p{7em}p{5em}p{7em}p{6em}p{5.5em}p{5em}}
    \toprule
    \textbf{Datasets} & \textbf{Method} & \textbf{Train data(days)} & \textbf{Precision (\%)} & \textbf{Recall (\%)}  & \textbf{ACC (\%)} \\

    \midrule
    \multirow{3}{*}{CADETS} 
    & Threatrace                           & 11  & 0.28   & 90.91     & 98.03     \\
    \cmidrule{2-6}
    & KAIRO                                & 3   & 61.11   & 100       & 99.99     \\
    \cmidrule{2-6}
    & CLIProv                              & 3   & 76.67   & 100      & 99.99    \\
    \midrule
    \multirow{3}{*}{CADETS(2-hop)}
    & Threatrace                           & 11  & 90.42   & 99.97      & 99.96     \\
    \cmidrule{2-6}
    & KAIRO  & 3   & 100     & 100        & 100      \\
    \cmidrule{2-6}
    & CLIProv & 3 & 100     & 100       & 100      \\
    \midrule
    \multirow{3}{*}{THEIA} 
    & Threatrace                           & 9   & 0.08   & 95.65     & 99.18    \\
    \cmidrule{2-6}
    & KAIRO                                & 3   & 75.00   & 100      & 99.99   \\
    \cmidrule{2-6}
    & CLIProv                              & 2   & 82.14   & 100       & 99.99  \\
    \midrule
    \multirow{3}{*}{THEIA(2-hop)} 
    & Threatrace                           & 11  & 90.42   & 99.97      & 99.96     \\
    \cmidrule{2-6}
    & KAIRO & 3   & 100     & 100       & 100     \\
    \cmidrule{2-6}
    & CLIProv  & 2 & 100     & 100         & 100      \\
    \bottomrule
    \end{tabular}
\end{table*}

Experimental results show that Threatrace's neighborhood aggregation mechanism marks a large number of neighbor nodes as anomalies. Specifically, ThreaTrace identified 12848 attack nodes in the CADETS dataset and 25297 attack nodes in the THEIA dataset. However, according to the DARPA attack report, the CADETS dataset contains only 69 attack nodes and the THEIA dataset contains only 71 attack nodes. Threatrace misidentifies a large number of neighbor nodes unrelated to attacks, such as system functional entities, as attack nodes, increasing the workload of security analysis and the cost of false positives. In contrast, the attack scenarios generated by KAIROS and CLIProv are more focused on the actual attack path. In the THEIA dataset browser extension attack scenario, CLIProv identified 24 attack nodes and 29 edges, and KAIROS identified 25 attack nodes and 31 edges, which effectively highlighted the key steps of the actual attack and helped security analysts quickly discover the attack chain. By jointly learning the behavior patterns of attack behavior and benign behavior, CLIProv alleviates the dependence of the model on the distribution of benign behavior, and can achieve the best detection performance with a small amount of training data, showing stronger generalization ability and practical application value.

\subsubsection{Ablation study}
\label{6_3_3}
To evaluate the impact of different components of CLIProv, we conducted two sets of ablation experiments for graph-level detections. For the multimodal feature representation method, we adopt a single encoder architecture and model the attack sequence identification as a binary classification task, using only the provenance log for training and prediction. By fine-tuning the log encoder, we evaluate the attack identification performance on single-modal data. For the attack detection task, we use a dual encoder architecture to simultaneously learn vector representations of threat intelligence and provenance logs. Through the joint training of contrastive loss and classification loss, the model strengthens the class discrimination ability while maintaining semantic alignment, and finally identifies the attack behavior through the binary classification task. The experimental results for different models are presented in Table~\ref{tab_7}.

\begin{table}[htbp]
  \centering
  \caption{Comparison experiments of different models}
  \label{tab_7}
    \begin{tabular}{p{4em}p{6em}p{3em}p{3em}p{3em}}
    \toprule
    \textbf{Datasets} & \textbf{Method} & \textbf{Precision (\%)} & \textbf{Recall (\%)}  & \textbf{ACC (\%)} \\
    \midrule
    \multirow{3}{*}{CADETS} 
    & Single-modal         & 44.44    & 100    & 99.92 \\
    \cmidrule{2-5}
    & Cont+classifier    & 33.33    & 100    & 99.87 \\
    \cmidrule{2-5}
    & CLIProv              & 100     & 100    & 100 \\
    \midrule
    \multirow{3}{*}{THEIA} 
    & Single-modal         & 50.00    & 100   & 99.89 \\
    \cmidrule{2-5}
    & Cont+classifier    & 36.36    & 100     & 99.77 \\
    \cmidrule{2-5}
    & CLIProv              & 100     & 100   & 100 \\
    \bottomrule
    \end{tabular}
\end{table}

The experimental results show that the precision of the single-modal pre-trained model is relatively low, primarily due to the severe data imbalance in the provenance logs. This imbalance skews the model’s decision threshold toward the majority class, increasing the probability of misclassifying negative samples as positive. The use of multimodal learning, involving both contrastive loss and classification loss, resulted in more false positives. This occurs because attack behaviors are diverse, and contrastive pre-training causes attack behavior representations to become more dispersed in the semantic space. This dispersion blurs the classification boundaries, leading to an increase in false positives. CLIProv utilizes a threat intelligence repository to generate query texts and measures the distance between these texts and log sequences for attack retrieval. This approach effectively adapts to diverse attack patterns, significantly mitigating the issue of data imbalance. It not only enhances precision in attack detection but also improves the model's flexibility and reliability in handling complex attack behaviors.

\subsection{Impact of key parameters}
\label{sec:6-4}
In the previous section, we evaluated CLIProv using a fixed set of hyperparameters. We independently vary each of them and report their impact on detection and runtime performance.

\subsubsection{Numbers of text augmentations (${{n}_{aug}}$)} 

\begin{figure*}[ht]
    \centering
    \subfloat[]{\label{fig:6a}
    \includegraphics[width=3in]{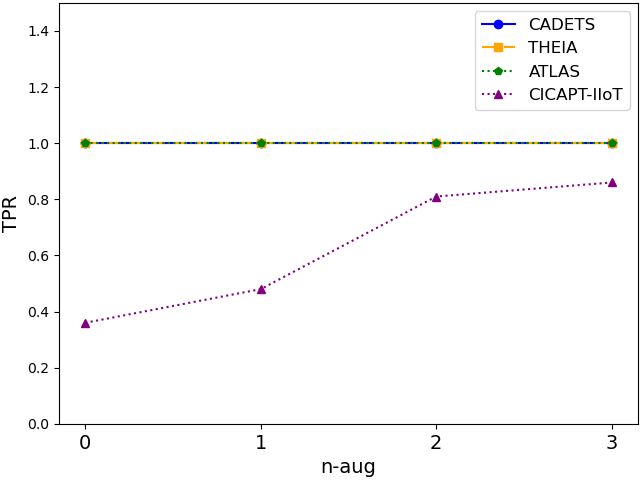}}
    \quad
    \subfloat[]{\label{fig:6b}
    \includegraphics[width=3in]{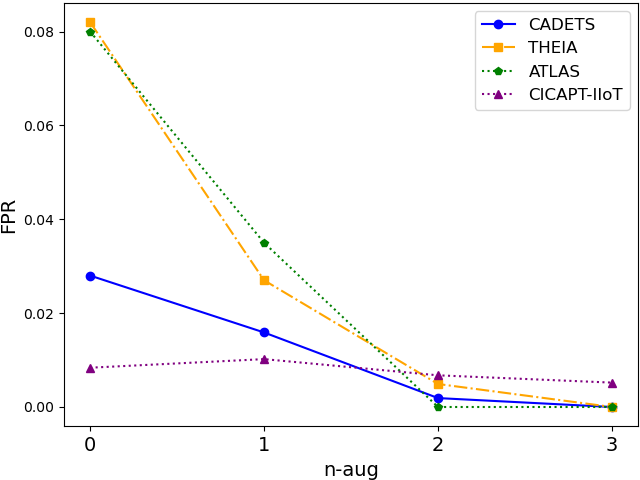}}
    \caption{TPR and FPR under different levels of data augmentation.}
    \label{fig_6}
\end{figure*}

Due to the limited number of attack samples in the provenance logs, there is a serious imbalance problem in the training data, which not only limits the generalization ability of the model, but also easily leads to false positives and false negatives. The data augmentation method based on GPT-3 enriched the semantic representation of attack behaviors by generating diverse attack descriptions, which helped the model learn a more fine-grained representation of attack behaviors in the semantic space, so as to better distinguish attacks from normal behaviors. As shown in Fig.~\ref{fig_6}, with the increase of the number of enhanced samples, the True Positive Rate (TPR) of CICAPT-IIoT dataset rises significantly, indicating that the ability of the model to identify attacks is enhanced and the generalization performance is improved. Additionally, the False Positive Rate (FPR) decreases across all datasets, reflecting that the augmented samples help the model more accurately model the boundary between attack and normal behavior.

\subsubsection{Sampling percentage of benign sequences in the training data} 

\begin{figure*}[ht]
    \centering
    \subfloat[]{\label{fig:7a}
    \includegraphics[width=3in]{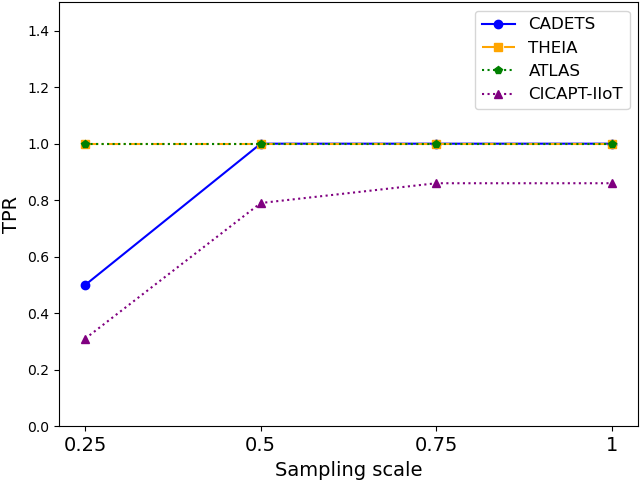}}
    \quad
    \subfloat[]{\label{fig:7b}
    \includegraphics[width=3in]{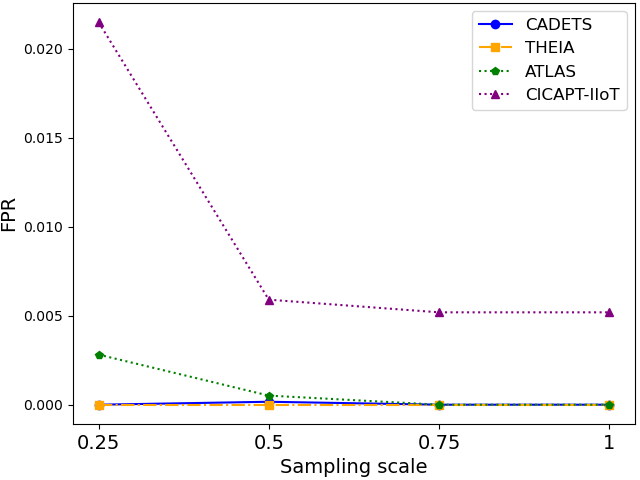}}
    \caption{Impact of benign sequence sampling scale on TPR and FPR.}
    \label{fig_7}
\end{figure*}

The CLIProv model simultaneously learns features from both attack and benign sequences, matching log sequences with threat intelligence without needing to learn from all benign sequences. We randomly sample the benign log sequence and take the average of the results of five experiments, as shown in Fig.~\ref{fig_7}. The results show that at a 25\% sampling rate, the TPR of the CADETS and CICAPT-IIoT datasets decreases, indicating insufficient learning of benign behaviors, which causes the model to misclassify attack sequences as benign. Additionally, with a low sampling rate of benign sequences, both the ATLAS and CICAPT-IIoT datasets exhibit a higher FPR. However, at a 50\% sampling rate, performance improves across all datasets. At a 75\% sampling rate, performance comparable to training with all benign sequences is achieved, with a reduced training time of 159 minutes than using all benign sequences.

\subsection{Interpretable attack scenarios}
\label{6_5}
\begin{figure}[!t]
\centering
\includegraphics[width=3.5in]{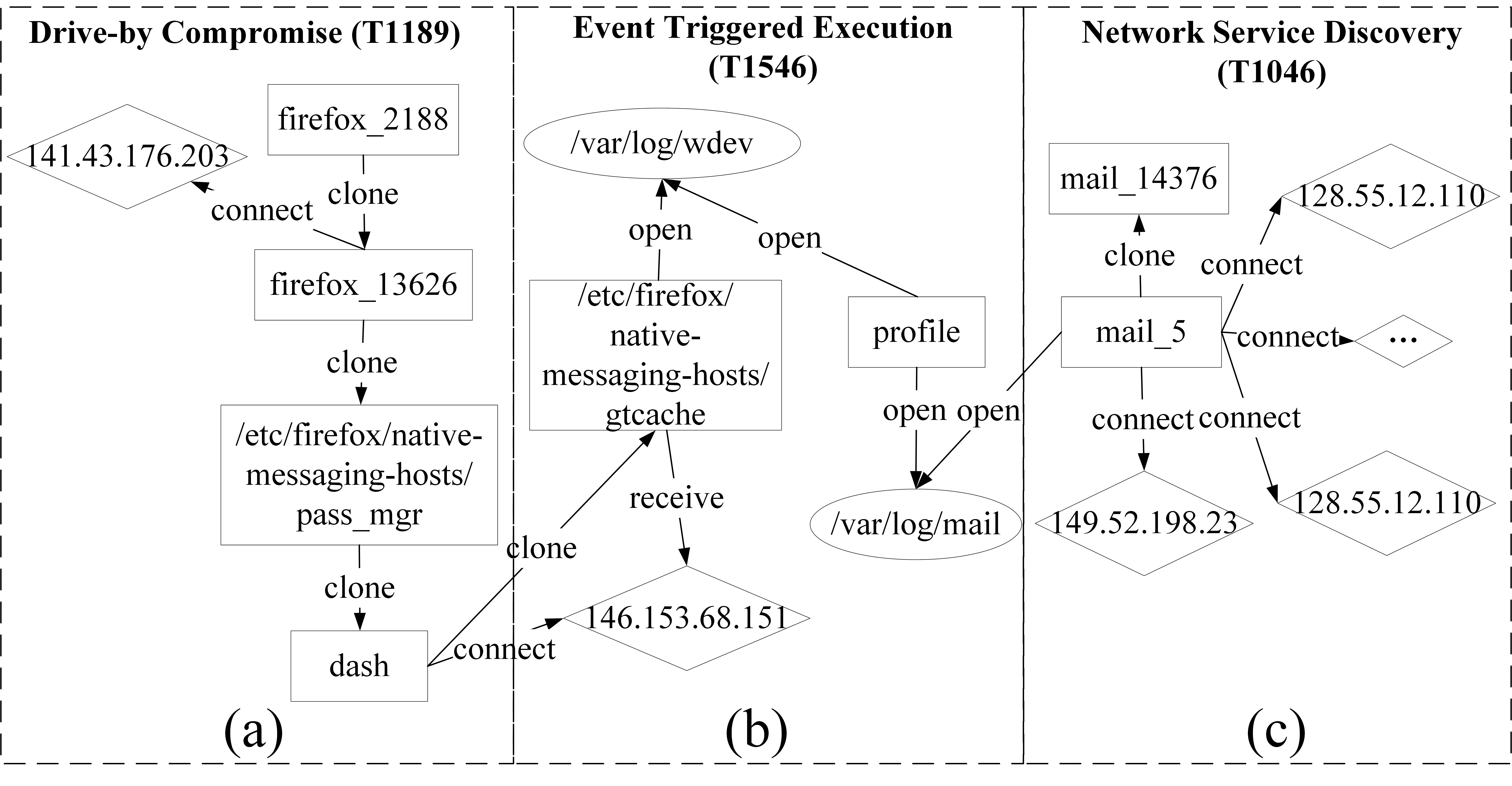}
\caption{An interpretable attack graphs example generated by CLIProv.}
\label{fig_9}
\end{figure}

In this section, we provide an example showing the interpretable attack graph generated by CLIProv, as shown in Fig.~\ref{fig_9}. Due to the large size of the original attack graphs, we manually simplified them to highlight only the critical attack steps.

The browser extension attack scenario for THEIA dataset consists of three system behavior subgraphs, each of which corresponds to a different threat intelligence: (a) Drive-by Compromise (T1189): \textit{``Mustard Tempest has used drive-by downloads for initial infection, often using fake browser updates as a lure.''} (b) Event Triggered Execution (T1546): \textit{``Green Lambert can establish persistence on a compromised host by modifying profile, login, and run command files associated with the bash, csh, and tcsh shells.''} (c) Network Service Discovery (T1046): \textit{``APT32 performed network scanning to search for open ports, services, OS fingerprinting, and other vulnerabilities.''} It can be found that there are similarities between the threat intelligence and the attack behavior patterns in the attack behavior subgraph. In the first phase, the attacker exploits the \textit{Firefox} vulnerability and connects to the external shellcode server\textit{ 141.43.176.203}, using Firefox's \textit{Native Messaging mechanism} to establish communication between the browser extension and the native application. In the second phase, \textit{pass\_mgr} clones and executes the \textit{dash} shell, and starts the malicious process \textit{gtcache} to tamper with the system log files \textit{var/log/wdev} to control the \textit{profile} for persistence. In the third stage, \textit{profile} injects malicious code into the \textit{mail} process, and uses mail to scan multiple ports of external IP to explore potential vulnerabilities or attack points.

\subsection{Runtime Performanc}

In this subsection, we evaluate the runtime overhead of each component and present the results in Table~\ref{tab_10}. The results are averaged over multiple attack scenarios across different datasets. Subsequently, we discuss the density of attack nodes within the attack scenarios generated by CLIProv.

\begin{table}[htbp]
\centering
\caption{Overhead of each component}
\label{tab_10}
\begin{tabular}{p{15em}p{7em}}
\toprule
\textbf{Component} & \textbf{Mean Duration} \\
\midrule
Preprocessing & 13.9~min \\
Representation learning & 8.72~h \\
Threat searching and investigation & 19.15~s \\
Each system behavior subgraph & 8.27~ms \\
\bottomrule
\end{tabular}
\end{table}

On average, CLIProv takes 13.9 minutes to process the logs of a single attack scenario from a given dataset. Training the multimodal model on data from 14 attack scenarios across three datasets requires 8.72 hours for 100 epochs. For threat searching and investigation, the query time per scenario averages 19.15 seconds, with each system behavior subgraph’s corresponding log sequence processed in 8.27 milliseconds. It's important to note that training was performed on a single GPU server. If distributed training were implemented, the overall overhead would likely be significantly reduced by leveraging parallel processing.

\section{Discussion and Limitation}
\label{sec:7}

\subsection{Limitations of knowledge} 

CLIProv achieves semantic alignment between provenance logs and threat intelligence through multimodal learning. By leveraging attack pattern information from threat intelligence, the model can recognize and understand hierarchical features within attack scenarios, ranging from individual system events to the broader intent behind a series of system behaviors. Compared to query graph-based methods, CLIProv does not require accurate query graphs and can interpret log behaviors through the semantic information provided by threat intelligence. However, as a knowledge-based threat hunting method, CLIProv is inherently limited by the scope of the attack knowledge it has been trained on. Although CLIProv demonstrates a certain degree of generalization, its performance may decrease when encountering attack patterns that it has not been trained on. For instance, attacks involving techniques such as Wi-Fi discovery, which are significantly different from those in the training set, may not be effectively detected. One potential solution is to train a more granular log encoder at the system event level. This approach could learn the semantics of log events to improve its adaptability to previously unseen attack techniques.

\subsection{Noise in the attack scenario} 

CLIProv leverages the spatiotemporal characteristics of APT attacks to partition provenance graphs, which can effectively isolate distinct subgraphs of system behaviors. However, during the attack process, it is inevitable to call some system service-related files, resulting in the generated attack scenarios containing nodes that are not related to the attack. To address this issue, we can consider integrating anomaly-based investigation methods. By training a new model to learn the behavior patterns of benign nodes, or by scoring the benignity of system nodes using frequency and statistical techniques, it is possible to further filter out irrelevant nodes from the attack scenarios.

\subsection{Fake threat intelligence} 

One potential risk of external intelligence-based threat hunting methods is fake intelligence, which could lead to data poisoning attacks on detection systems. An informed adversary might manipulate the system by injecting false threat intelligence \cite{ranade2021generating,song2023generating}, causing the model to learn incorrect inputs that align with the attacker's malicious intent. To mitigate this risk, Mahlangu \cite{mahlangu2019data} ensures the credibility of the intelligence database by validating the sources of intelligence, while Mitra et al. \cite{mitra2021combating} propose generating cybersecurity knowledge graphs from provenance graphs, calculating quality scores for entities and relationships to assess their trustworthiness. In future training, incorporating fake intelligence samples as negative samples for adversarial training could enhance the model's robustness against false intelligence attacks.

\section{Conclusion}
\label{sec:8}
In this paper, we present CLIProv, a knowledge-based system that identifies APT attacks from provenance logs and generates complete attack scenarios. CLIProv innovatively aligns log sequences with threat intelligence through multimodal learning, enabling semantic search for attack behaviors in provenance logs. We evaluate CLIProv across four datasets, demonstrating its high precision in threat identification with minimal time overhead. The system generates interpretable attack scenarios, offering valuable insights for analysts in their investigations. In future work, we aim to develop an event-level log pre-training model that incorporates semantic understanding and contextual relationships at the event level. This model will enhance the system’s ability to accurately capture the associations and causal relationships between log events, improving its adaptability to new attack methods and techniques through fine-grained semantic learning.







\printcredits

\bibliographystyle{model1-num-names}
\bibliography{cas-refs}

\end{document}